# Synthesis of sub-5 nm Co-doped $SnO_2$ nanoparticles and their structural, microstructural, optical and photocatalytic properties

T. Entradas[1], J. Cabrita[1], S. Dalui[2], M.R. Nunes[1], O.C. Monteiro[1*] and A.J. Silvestre[3*]

[1]Department of Chemistry and Biochemistry and CQB, Faculty of Sciences, University of Lisbon, Campo Grande, 1749-016 Lisboa, Portugal.

[2]Department of Physics and ICEMS, Faculty of Sciences, University of Lisbon, Campo Grande, 1749-016 Lisboa, Portugal.

[3]Department of Physics and ICEMS, Instituto Superior de Engenharia de Lisboa - ISEL, R. Conselheiro Emídio Navarro 1, 1959-007 Lisboa, Portugal.

## ABSTRACT

A swift chemical route to synthesize Co-doped $SnO_2$ nanopowders is described. Pure and highly stable $Sn_{1-x}Co_xO_{2-\delta}$ ($0 \leq x \leq 0.15$) crystalline nanoparticles were synthesized, with mean grain sizes < 5 nm and the dopant element homogeneously distributed in substitutional sites of the $SnO_2$ matrix. The UV-visible diffuse reflectance spectra of the $Sn_{1-x}Co_xO_{2-\delta}$ samples reveal red shifts, the optical bandgap energies decreasing with increasing Co concentration. The samples' Urbach energies were calculated and correlated with their bandgap energies. The photocatalytic activity of the $Sn_{1-x}Co_xO_{2-\delta}$ samples was investigated for the 4-hydroxylbenzoic acid (4-HBA) degradation process. A complete photodegradation of a 10 ppm 4-HBA solution was achieved using 0.02% (w/w) of $Sn_{0.95}Co_{0.05}O_{2-\delta}$ nanoparticles in 60 min of irradiation.

---

*Authors to whom correspondence should be addressed: Telephone: +351-217500865, Fax: +351-217500088, Email: ocmonteiro@fc.ul.pt (O.C. Monteiro); Telephone: +351-218317097, Fax: +351-21837138, Email: asilvestre@deq.isel.ipl.pt (A.J. Silvestre).

## 1. INTRODUCTION

Tin dioxide ($SnO_2$) is a wide bandgap metal oxide semiconductor in which inherent oxygen vacancies act as *n*-type dopants [1,2]. It has a *rutile*-type tetragonal structure based on octahedra wherein each tin atom is surrounded by six oxygen atoms, and its space group is $P_{42/mnm}$. The standard lattice parameters are $a = b = 4.7382$ Å and $c = 3.1871$ Å [3]. At room temperature $SnO_2$ has an optical bandgap of ~3.6 eV [4,5]. From the applications point of view, $SnO_2$ has been investigated in a broad variety of fields, including gas, *pH* and bio-sensors [6-11], transparent conducting electrodes [12,13], lithium ion batteries [14,15], field effect transistors [16], sensitized solar cells [17,18] and varistors [19]. $SnO_2$ has also been used in catalysis, as metal oxide support for precious metal catalysts in systems such as $Pt/SnO_2$, $Pd/SnO_2$, $Ru/SnO_2$ and $Rh/SnO_2$ [20], for the combustion at low temperatures of volatile organic compounds. Moreover, Co-doped $SnO_2$ has gained immense interest, being one of the first oxide-based diluted magnetic semiconductor (DMS) systems to be investigated [21], in parallel with the $Co:TiO_2$ [22,23] and Co:ZnO [24] systems, exhibiting ferromagnetic order well above room temperature.

The use of $SnO_2$ in photocatalysis has been much less studied, despite its crystalline structure being very similar to that of $TiO_2$, a worldwide used photocatalyst [25]. Indeed, a crystal network of corner-sharing octahedral units is considered as a pre-requisite for high photocatalytic activity since such a network increases the mobility of electrons and holes and consequently affects the probability of electrons and holes reaching the reaction sites on the surface of the photocatalyst [26]. Besides the crystal structure, a good photocatalyst should also have an adequate absorption band and a high specific surface area. The former property can be achieved, for instance, by tailoring the optical bandgap through doping [27], and the latter by controlling the particle size. However, it is still a challenge to prepare semiconductor nanocrystals with controllable size, shape and doping. In particular, the synthesis of nanoparticles of sizes under 10 nm still remains a major challenge [28]. They usually require complex synthesis procedures and/or suffer problems of aggregation or poor monodispersity, which strongly influence their properties and restrict their large-scale industrial production.

This work reports on a swift chemical route to synthesise highly stable undoped and Co-doped $SnO_2$ nanoparticles, with different dopant concentrations and grain sizes bellow 5 nm. The synthesis procedure is highly reproducible, low-cost and easily scaled-up. The influence of the Co doping concentration on the structure, microstructure and optical properties of the different synthesized samples was studied. The potential applicability of these materials for organic pollutant remediation processes was investigated, by studying their photocatalytic performance in the 4-hydroxylbenzoic acid (4-HBA) degradation reaction under UV-Vis irradiation.

## 2. EXPERIMENTAL

### *2.1. Materials*

All reagents were of analytical grade and were used as received without further purification. The solutions were prepared with bi-distilled water.

### *2.2. Nanoparticles synthesis*

A flowchart of the overall synthesis method is shown in Figure 1. A tin tetrachloride solution (Aldrich) diluted in a ratio of 2:5 in standard hydrochloride acid (37% HCl, Panreac), following a dilution up to a 0.1 M concentration was used as tin source. To this solution an ammonium 4 M solution (Merck) was added dropwise under vigorous stirring, until a complete precipitation of a white solid was observed. The resulting suspension was kept at rest for 15 hours at room temperature, and then filtered and vigorously rinsed with deionized water in order to remove the remaining ammonium and chloride ions. Crystallization of the $SnO_2$ precursor was done in autoclave at 200 ºC, for 12 hours in aqueous suspension. The same chemical route was used to synthesize Co-doped $SnO_2$, by adding the required molar amount of metallic cobalt (Johnson Matthey) solubilized in nitric acid to the tin tetrachloride solution (doping step). After being washed, the undoped as well as the Co-doped $SnO_2$ nanoparticles can be easily stored in aqueous suspensions and straightforwardly retrieved by centrifugation whenever necessary. For this study, $Sn_{1-x}Co_xO_{2-\delta}$ samples with nominal Co/Sn ratios of $x$ = 0.005, 0.01, 0.03, 0.05, 0.10 and 0.15 were prepared.

### *2.3. Photodegradation experiments*

Adsorption studies were carried out using $Sn_{1-x}Co_xO_{2-\delta}$ nanoparticle suspensions in 4-HBA aqueous solution (10 ppm) under stirring in dark conditions. After centrifugation, the 4-HBA concentration was estimated by measuring the absorbance of its characteristic band centred at 248 nm.

The photodegradation experiments were conducted using a 250 ml refrigerated photo-reactor [29]. A 450 W Hanovia medium-pressure mercury-vapour lamp was used as radiation source, the total irradiated energy is 40-48% in the ultraviolet range and 40-43% in the visible region of the electromagnetic spectrum. Suspensions were prepared by adding 30 mg of powder to 150 mL of a 10 ppm 4-HBA aqueous solution. Prior to irradiation, suspensions were stirred in darkness for 1 hour to ensure adsorption equilibrium. During irradiation, suspensions were sampled at regular intervals, centrifuged and analyzed by UV-vis spectroscopy.

### *2.4. Characterization*

X-ray powder diffraction was performed using a Philips X-ray diffractometer (PW 1730) with automatic data acquisition (APD Philips v3.6B), using Cu Kα radiation ($\lambda$ = 0.15406 nm) and working at 40 kV/30 mA. The diffraction patterns were collected in the $2\theta$ range of 20º - 60º with a 0.02º step size and an acquisition time of 2.0 s/step. The diffractometer was calibrated before every measurement, and the instrumental broadening ($1.61\times10^{-3}$ rad) was measured from a standard macrocrystalline and strain-free silicon sample. The $K\alpha_2$ contribution was removed before the XRD pattern analyses. The $2\theta$ angular position of the diffraction peaks and their full-width at half-maximum, $\beta$, were calculated by fitting the experimental diffraction lines with a Pseudo-Voigt function. The $\beta$ values were corrected taking into account the instrumental broadening. Transmission electron microscopy (TEM) and selected area electron diffraction (SAED) were carried out using a JEOL 200CX microscope operating at 200 kV. The elemental composition of the samples was analysed by energy dispersive X-ray spectroscopy (EDS). UV–VIS absorption spectra of the solutions were recorded using a UV–VIS absorption spectrophotometer (Shimadzu, UV-2600PC). The powder diffuse reflectance spectra (DRS) were recorded in the wavelength range of 220-1400 nm using the same equipment with an ISR 2600plus integration sphere.

## 3. RESULTS AND DISCUSSION

### *3.1. Structure and morphology*

Figure 2 shows a representative set of diffractograms recorded for the synthesized undoped $SnO_2$ and $Sn_{1-x}Co_xO_{2-\delta}$ nanopowder samples. The diffraction reflections were indexed on the basis of the tetragonal rutile $SnO_2$ phase, using the JCPDS database card no. 41-1445 [3]. All the XRD patterns show broad peaks matching the expected diffraction reflections of the (110), (101), (200), (211) and (220) rutile $SnO_2$ planes, with similar relative intensities. No traces of other undesirable phases were observed throughout the whole range of Co contents considered, *e.g.* cobalt clusters, cobalt oxides, Co-Sn or Co-Sn-O phases, even when the diffracted intensity was plotted on a logarithmic scale. Note, however, that the presence of Co in addition to Sn and O was confirmed for all Co-doped samples by EDS, with a Co:Sn ratio close to the nominal stoichiometry (not shown). Therefore, the XRD patterns seem to provide evidence for the homogeneous distribution of Co into the $SnO_2$ matrix, within the detection limit of the technique.

The effect of Co-doping on the structure of $SnO_2$ was studied by analysing the lattice parameters, the unit cell volume and the microstrain of the powder samples. Based on the interplanar distance, $d_{hkl}$, between the ($hkl$) planes of a tetragonal system, the lattice parameters of the different synthesized samples were calculated using the relations

$$\begin{cases} a = \sqrt{2}\, d_{110} \\ c = \dfrac{a\, d_{101}}{\sqrt{a^2 - d_{101}^2}} \end{cases}, \tag{1}$$

taking into account Bragg's law $d_{hkl} = \lambda / 2\sin\theta_{hkl}$ and the $2\theta$ angular positions of the (110) and (101) peaks in the XRD patterns (Figure 2). The lattice parameters obtained as well as the unit cell volume ($V = a^2 c$) are given in Table 1. While the $a$ parameter increases almost linearly with increasing Co content, the $c$ parameter has a maximum for $x = 0.03$. Nonetheless, the overall unit cell volume increases by about 0.34% when $x$ increases from 0 to 0.15, as can be seen in Figure 3. This variation is related with the ionic radius size effect of $Co^{2+}$ ($r = 0.745$ Å) when replacing $Sn^{4+}$ ($r = 0.69$ Å) [30] into the $SnO_6$ octahedra $SnO_2$ building blocks, and consequent increase of oxygen vacancies due to the unbalanced electrical charge, which also causes cation repulsion and a unit cell volume increase. The samples' microstrain, $\varepsilon$, was evaluated using the formula

$$\varepsilon = \frac{d_{hkl} - d_{hkl}^{s}}{d_{hkl}^{s}}, \tag{2}$$

were $d_{hkl}^{s}$ stands for the standard interplanar distance between the ($hkl$) planes. The $\varepsilon$ values were estimated using the interplanar distance between the (110) planes, and are given in Table 1. As can be seen, the microstrain of all samples are positive (tensile strain), increasing from $0.011\times10^{-3}$ to $1.759\times10^{-3}$ with increasing Co content.

In what concerns the samples' mean crystallite sizes, <$D$>, they were evaluated by Scherrer's equation [31] using the (110) reflections. The diffractograms of the different samples enabled to estimate <$D$> between 3.25 nm and 3.42 nm. These values are similar, and no correlation between the mean crystallite size and the samples' Co content could be deduced (see Table 1).

Given the above mentioned mean crystallites sizes and assuming spherically shaped particles, the number of unit cells per particle is only about 250.

The morphology and structure of the nanoparticle samples were further investigated by transmission electron microscopy. Figure 4 shows bright field TEM micrographs of a representative set of synthesised nanopowder samples as well as SAED images and the respective particle size histograms. As shown, the undoped $SnO_2$ sample is composed of nanoparticles showing a *quasi*-spherical shape, which can be made almost monodisperse by acidifying the particles' suspension with nitric acid (inset upper-left TEM micrograph). This result could probably be attributed to double layer repulsion phenomena, well known to be responsible for the stability of colloidal systems [32]. Since the sample consists of a very large number of small randomly distributed crystals, the SAED pattern consist of well-defined continuous diffraction rings, which can be indexed, from inside to outside, to the (110), (101), (200) and (211) planes of rutile $SnO_2$, in accordance with XRD reflections described above. The sample's particle size follows a lognormal distribution with mean size 3.82±1.06 nm, in good agreement with the size of the coherent diffracting domains calculated from the XRD pattern using Scherrer's equation. Analogous microstructures were found for the Co-doped samples. SAED images of each Co-doped samples are shown as insets in the corresponding TEM micrograph. The analysis of the TEM images show that Co-doped samples present similar spheroidal morphologies as found for the undoped samples, with particle sizes also following a lognormal distribution with mean values varying between 3.59±0.71 nm and 4.06±0.96 (see Table 1), in accordance with the $<D>$ values found using XRD data. The SAED patterns of the $Sn_{1-x}Co_xO_{2-\delta}$ samples are comparable to that described above for the undoped sample.

### *3.2. UV-Vis spectroscopy*

The optical characterization of the undoped and Co-doped $SnO_2$ samples was carried out by measuring their diffuse reflectance, $R$, at room temperature. $R$ can be related to the absorption Kubelka-Munk function $F_{KM}$, by the relation $F_{KM}(R) = (1-R)^2/2R$, which is proportional to the absorption coefficient [33]. The obtained absorption spectra are shown in Figure 5. A significant red shift in the optical absorption band edge can be seen for the Co-doped samples compared to the undoped sample. The optical bandgap energies of the samples, $E_g$, were estimated by plotting the function $f_{KM} = (F_{KM}h\nu)^2$ *vs.* $h\nu$ (Tauc plot), where $h$ stands for Planck's constant and $\nu$ for the radiation frequency, and by extrapolating the linear portion of the curve to zero absorption (Figure 6). The different extrapolated $E_g$ values are given in Table 2. The optical bandgap of the undoped $SnO_2$ sample was estimated as 3.72 ± 0.03 eV which is blue shifted by 0.12 eV relative to the $SnO_2$ bulk value (3.6 eV). One may wonder if such blue shift might originate from the quantum confinement phenomenon associated with the nanosized crystallites that form the synthesized samples [34]. The structural size effect can be important if the crystallite size is in the range of the Bohr radius of the first excitonic state, which is given by [35]

$$r_B = \frac{m_0\, \varepsilon_r}{\mu} a_B \,, \tag{3}$$

where $m_0$ stands for the electron mass, $\varepsilon_r$ for the relative dielectric permittivity, $\mu$ for the effective reduced electron-hole mass, and $a_B$ for the Bohr radius of the hydrogen atom ($5.292\times10^{-11}$ m). For the $SnO_2$, $\varepsilon_r$ = 14, the electron effective mass is $m_e^* = 0.275\, m_0$ and the hole effective mass is $m_h^* \ll m_e^*$ (*i.e.* $\mu \simeq 0.275\, m_0$) [36]. Computing these values in equation 3, we get $r_B$ = 2.7 nm. Since $2r_B$ is greater than the average crystallite size of the undoped $SnO_2$ nanopowder sample (see Table 2), its optical behaviour can readily be ascribed to the effect of particle size. Note however that the $E_g$ blue shift of this sample (0.12 eV) is lower than that we might expect (0.52 eV) using Brus' model [34]. The opposite behaviour was found for the optical bandgap energies of Co-doped $SnO_2$ samples. All doped samples present $E_g$ values red shifted relative to the standard $SnO_2$ bulk value, despite having nanoparticle sizes similar to that determine for the undoped $SnO_2$ sample. Figure 7 shows the $E_g$ values as a function of the Co content. As can be seen, the optical bandgap energies of the $Sn_{1-x}Co_xO_{2-\delta}$ samples decrease with increasing Co content, the $E_g$ values sharply decreasing to 3.23 ± 0.08 eV for $x$ = 0.03 before slowing down when approaching the lower limit of 3.19 ± 0.09 eV for $x$ = 0.15. Similar narrowing of the optical bandgap has been reported in the literature for Co-doped $SnO_2$ [37-39] as well as for Fe-doped $SnO_2$ [40,41] nanopowders. Since the synthesis procedure for the Co-doped $SnO_2$ samples was similar to that used to synthesize the undoped samples, it can be concluded that the red shift deduced for the optical bandgap of the doped samples results from the Co doping process. Furthermore, this result seems to be consistent with the hypothesis that Co is homogeneously distributed in the substitutional sites of the $SnO_6$ octahedra tin oxide based structures, and thus gives rise to $Sn_{1-x}Co_xO_{2-\delta}$ doped structures with oxygen deficiency due to the lower valence states of cobalt compared to tin, as previously mentioned. Indeed, the red shifts of Co-doped $SnO_2$ are consistent with the introduction of electronic states in the tin oxide bandgap associated with the 3*d* electrons of $Co^{2+}$ cations and oxygen defects, in a similar way to what occurs in Co-doped $TiO_2$ [42], and for which band-structure calculations have shown that the valence band derives primarily from O 2*p*-levels, the conduction band from the Ti 3*d*-levels, and that the crystal-field split Co 3*d*-levels form localized bands within the original bandgap of $TiO_2$ [43-45]. The results reported here seem to support that comparable effects may occur when Co replaces Sn in the $SnO_2$ host lattice. The localized states induced by Co doping are expected to form tails of states that extend the bands into the bandgap, producing an absorption tail known as Urbach tail [46]. The Urbach energy, $E_U$, associated with the width of the Urbach tail obeys the exponential law [46]

$$\alpha = \alpha_0 \, e^{\frac{h\nu}{E_U}} , \qquad (4)$$

where $\alpha$ is the optical absorption coefficient and $\alpha_0$ is a constant. Since the absorption Kubelka-Munk function $F_{KM}$ is proportional to the sample's absorption, the Urbach energies of the different synthesized samples were estimated from the slopes of $\ln(F_{KM})$ plotted as a function of photon energy, in an energy range just below the gap (not shown). The calculated $E_U$ values as a function of the samples' Co content are plotted in Figure 8 and given in Table 2. As shown, $E_U$ varies from 0.179 eV to 1.749 eV as $x$ increases from 0 to 0.15, the higher $E_U$ values indicating further introduction of tails into the bandgap of the $Sn_{1-x}Co_xO_{2-\delta}$ samples. This can be further highlighted by plotting $E_g$ as a function of $E_U$, as shown in the inset of Figure 8. As the Urbach energy increases, the optical bandgap energy decreases, increasing the red shift due to possible band-to-tail and tail-to-tail transitions. Moreover, the optical bandgap decreases as the Urbach energy increases, with a fairly linear dependence, which reflects the influence of the structural disorder increase on the width of the absorption tail of the synthesized nanopowders as the Co content increases [47,48], in accordance with the increase of the samples' microstrain as previously discussed. Studies of the Urbach energies of $SnO_2$ are scarce in literature. A value of 0.037 eV was reported for single crystals of $SnO_2$ [49]. However, higher Urbach energies have already been found either for films or for particular $SnO_2$ based materials, mostly with amorphous or nanocrystalline phases. Melsheimer and Ziegler [50] studied the Urbach tail of $SnO_2$ thin films deposited by the spray pyrolysis technique, reporting $E_U$ values of 0.220 eV for films deposited at 500 ºC (polycrystalline films) and 0.530 eV for films deposited at 340 ºC (amorphous films). More recently Chetri and Choudhury [51] reported Urbach energies of 0.424 and 0.432 eV for $SnO_2$ nanoparticles annealed at 200 ºC (crystallite size 2.53 nm) and 600 ºC (crystallite size 6.45 nm), respectively. Regarding Co-doped $SnO_2$, Habubi *et al.* [52] reported $E_U$ values of 0.512 eV and 0.588 eV for Co:$SnO_2$ films with 3% and 7% of cobalt, respectively, and crystallite sizes of about 50 nm.

### *3.3. 4-HBA photodegradation*

The photocatalytic activity of the $Sn_{1-x}Co_xO_{2-\delta}$ nanopowders was studied using 4-HBA as model pollutant. Prior to photocatalysis, the adsorption ability of the samples was tested and it was found that after one hour in dark conditions no 4-HBA adsorption was noticed, either on the $SnO_2$ or on the Co doped-$SnO_2$ surfaces. The 4-HBA typical absorption spectrum is characterized by one broad absorption band centred at 248 nm, which is related with the absorption by the aromatic ring. This absorption peak was used as a reference for the photodegradation analysis.

Figure 9 shows the absorption spectra of a 10 ppm 4-HBA solution during irradiation in the presence of $Sn_{0.95}Co_{0.05}O_2$. A clear decrease of the 4-HBA characteristic 248 nm absorbance band intensity was observed with increasing irradiation time. Identical profile spectra were obtained for all the tested samples and for photolysis (not shown), which suggests an identical mechanism for the 4-HBA photocatalytic degradation process. A close inspection of the obtained absorption spectra (inset of Figure 9) shows, for early irradiation times, a decrease and a slight broadening of the 248 nm absorbance band. A simultaneous increase of the 270 - 315 nm absorption range arises. The broadening of the 248 nm absorbance band can be associated with the presence of benzoquinone (BQ), phenol, catechol (COH) and resorcinol (ROH), with maximum absorption peaks at 246, 270, 275 and 273 nm, respectively [53]. On the other hand the increase of the absorbance at 288 nm can be attributed to the presence of hydroquinone (HQ). This behaviour has been attributed to an initial and fast formation of the photodegradation's main byproducts (*e.g.* catechol, resorcinol, hydroquinone and benzoquinone) which absorb in the same 4-HBA wavelength range [53].

Figure 10 shows the 4-HBA absorption spectra obtained after 60 minutes of irradiation in the presence of the different $SnO_2$ based samples. As shown, either the undoped $SnO_2$ or the Co-doped $SnO_2$ powders have demonstrated photocatalytic activity on the 4-HBA degradation process. The best performance was obtained using nanoparticles with composition $Sn_{0.95}Co_{0.05}O_{2-\delta}$. Indeed, after 60 min of irradiation no peaks related either to 4-HBA or to its degradation byproducts appear in the final solution absorption spectrum. On the other hand, the sample with the lower photocatalytic ability for the 4-HBA degradation was the undoped $SnO_2$. Increasing Co doping up to 5%, a gradual improvement of the photocatalytic performance of $SnO_2$ was observed. Note, however, that for the sample with the highest Co-doping nominal amount (15%) a slight decrease of the photocatalytic activity resulted. This result suggests the existence of Co doping limit < 15% in order to optimize the photocatalytic activity of the $Sn_{1-x}Co_xO_{2-\delta}$ nanoparticles for the 4-HBA degradation.

## 4. CONCLUSIONS

A new and fast chemical route to synthesize single-phase Co-doped $SnO_2$ nanopowders was developed. It was shown that pure and highly stable $Sn_{1-x}Co_xO_{2-\delta}$ ($0 \leq x \leq 0.15$) nanoparticles can be synthesized under low temperature conditions, with mean grain sizes of about 3.5 nm and the dopant element homogeneously distributed in substitutional sites of the $SnO_2$ matrix. The microstrain of all samples are positive, increasing from $0.011\times10^{-3}$ to $1.759\times10^{-3}$ with increasing Co content. Consistently with *n*-type doping, the UV-visible diffuse reflectance spectra of representative powder samples have revealed red shifts, the optical bandgap energies decreasing with increasing Co concentration. The samples' Urbach energies were calculated and correlated with their bandgap energies. It was shown that the optical bandgap energy decreases, increasing the red shift due to possible band-to-tail and tail-to-tail transitions. Moreover, $E_g$ linearly decreases with $E_U$, which reflects the influence of the increasing structural disorder on the width of the absorption tail of the synthesized nanopowders as the Co content increases, in accordance with the samples' microstrain variation.

The photocatalytic behaviour of the $Sn_{1-x}Co_xO_{2-\delta}$ nanopowders was investigated for the 4-HBA degradation process. The best 4-HBA photocatalyst was $Sn_{0.95}Co_{0.05}O_{2-\delta}$. The complete photodegradation of a 10 ppm 4-HBA solution was achieved in 60 min, using 0.02% (w/w) of that material. Enhancement of the $SnO_2$ light absorption associated with an electron trapping phenomenon, which is a consequence of the Co-doping, is the most probable mechanism to explain these results.

**Acknowledgements**

This work was supported by Fundação para a Ciência e Tecnologia (FCT) under project No. PTCD/CTM/101033/2008. O.C. Monteiro acknowledges PEst-OE/QUI/UI0612/2013. The authors thank P.I.C. Teixeira for critically reading of the manuscript.

## Table captions

**Table 1** – Lattice parameters, unit cell volume, microstrain and crystallite size of the different nanopowders synthesized.

**Table 2** – Optical bandgap energies and Urbach energies of the different nanopowders synthesized.

**Figure captions**

**Figure 1** – Flowchart of the synthesis method developed to prepare either $SnO_2$ or Co-doped $SnO_2$ nanoparticles.

**Figure 2** – X-ray diffraction patterns of the undoped $SnO_2$ and $Sn_{1-x}Co_xO_{2-\delta}$ nanopowder samples.

**Figure 3** – Unit cell volume of the $Sn_{1-x}Co_xO_{2-\delta}$ nanopowders as a function of the samples' Co content. Note that the unit cell volume increases by only 0.34% when the Co/Sn ratio increases by 15%.

**Figure 4** – TEM micrographs of undoped $SnO_2$ and $Sn_{1-x}Co_xO_{2-\delta}$ nanopowder samples with $x = 0.05$, 0.10 and 0.15, and the corresponding particle size distribution histograms. The inset at the upper-right corner of each TEM micrograph shows the corresponding SAED image.

**Figure 5** – Diffuse reflectance spectra of the undoped $SnO_2$ and $Sn_{1-x}Co_xO_{2-\delta}$ nanopowder samples.

**Figure 6** – Tauc plots of the undoped $SnO_2$ and different $Sn_{1-x}Co_xO_{2-\delta}$ samples. The optical bandgap energies were estimated by extrapolating the linear portion of the curve to zero absorption.

**Figure 7** – Samples' optical bandgap energy *vs.* Co content. The dashed line is just a guide for the eye.

**Figure 8** – Urbach energy *vs.* samples' Co content. The dashed line is just a guide for the eye. The inset shows the samples' optical bandgap energies as a function of their Urbach energy.

**Figure 9** – Absorption spectra of a 4-HBA 10 ppm solution using 30 mg of $Sn_{0.995}Co_{0.005}O_{2-\delta}$ powder as photocatalyst, for different irradiation times. The inset shows a detail of the spectra in the 200 – 340 nm range. The peak positions of the main 4-HBA photodegradation byproducts are also shown (vertical dotted lines).

**Figure 10** – Absorption spectra of a 10 ppm 4-HBA solution using different powder samples as photocatalyst, for 60 minutes of irradiation: a) photolysis, b) $SnO_2$, c) $Sn_{0.995}Co_{0.005}O_{2-\delta}$, d) $Sn_{0.99}Co_{0.01}O_{2-\delta}$, e) $Sn_{0.97}Co_{0.03}O_{2-\delta}$, f) $Sn_{0.95}Co_{0.05}O_{2-\delta}$ and g) $Sn_{0.85}Co_{0.15}O_{2-\delta}$.

**Table 1**

| Sample | Nominal composition | Lattice parameters | | | Microstrain[a] ($\times 10^{-3}$) | Crystallite size (nm) | |
|---|---|---|---|---|---|---|---|
| | | ***a*** **(Å)** | ***c*** **(Å)** | **Cell volume (Å$^3$)** | | **Scherrer[b]** | **TEM[c]** |
| A | $SnO_2$ | 4.7334 | 3.1861 | 71.3849 | 0.011 | 3.25 | 3.82±1.06 |
| B | $Sn_{0.995}Co_{0.005}O_{2-\delta}$ | 4.7356 | 3.1879 | 71.4898 | 0.465 | 3.42 | 3.59±0.71 |
| C | $Sn_{0.99}Co_{0.01}O_{2-\delta}$ | 4.7361 | 3.1896 | 71.5467 | 0.587 | 3.30 | 3.65±0.77 |
| D | $Sn_{0.97}Co_{0.03}O_{2-\delta}$ | 4.7369 | 3.1899 | 71.5757 | 0.749 | 3.27 | 3.91±0.73 |
| E | $Sn_{0.95}Co_{0.05}O_{2-\delta}$ | 4.7395 | 3.1871 | 71.5905 | 1.289 | 3.18 | 4.04±1.06 |
| F | $Sn_{0.90}Co_{0.10}O_{2-\delta}$ | 4.7391 | 3.1892 | 71.6272 | 1.211 | 3.34 | 4.06±0.96 |
| G | $Sn_{0.85}Co_{0.15}O_{2-\delta}$ | 4.7417 | 3.1857 | 71.6270 | 1.759 | 3.21 | 3.98±0.90 |

[a] Values calculated using the interplanar distance between the (110) planes.
[b] Values calculated from XRD data using Scherrer equation.
[c] Values calculated from TEM image analyses.

**Table 2**

| Sample | Nominal composition | $E_g$ (eV) | $E_U$ (eV) |
|---|---|---|---|
| A | $SnO_2$ | 3.72±0.03 | 0.179±0.001 |
| B | $Sn_{0.995}Co_{0.005}O_{2-\delta}$ | 3.44±0.07 | 0.555±0.014 |
| C | $Sn_{0.99}Co_{0.01}O_{2-\delta}$ | 3.39±0.03 | 0.718±0.006 |
| D | $Sn_{0.97}Co_{0.03}O_{2-\delta}$ | 3.21±0.03 | 1.226±0.049 |
| E | $Sn_{0.95}Co_{0.05}O_{2-\delta}$ | 3.23±0.08 | 1.659±0.057 |
| F | $Sn_{0.90}Co_{0.10}O_{2-\delta}$ | 3.21±0.08 | 1.678±0.057 |
| G | $Sn_{0.85}Co_{0.15}O_{2-\delta}$ | 3.19±0.09 | 1.749±0.037 |

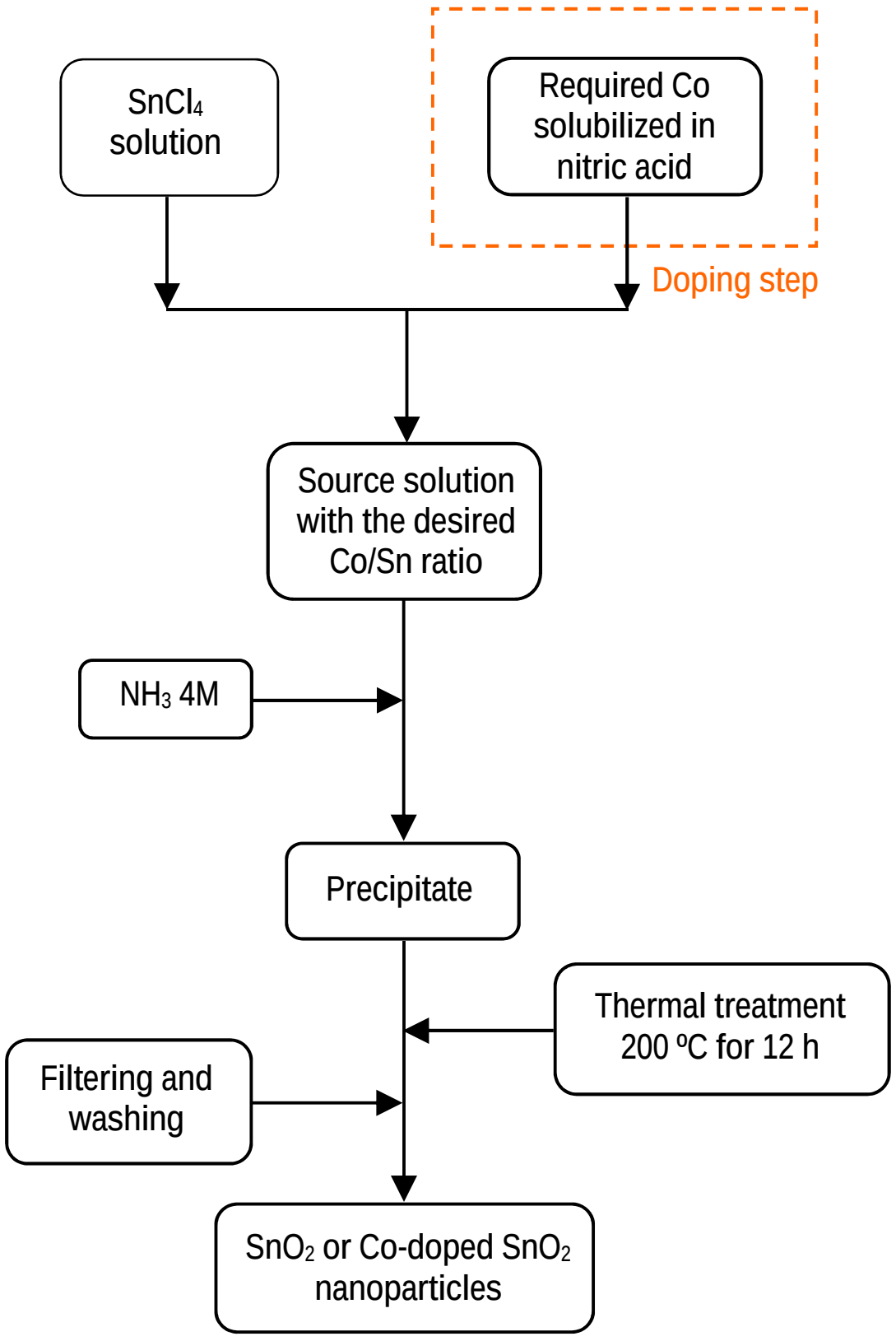

SnCl$_4$ solution
Required Co solubilized in nitric acid
Doping step
Source solution with the desired Co/Sn ratio
NH$_3$ 4M
Precipitate
Thermal treatment 200 ºC for 12 h
Filtering and washing
SnO$_2$ or Co-doped SnO$_2$ nanoparticles

**Figure 1**

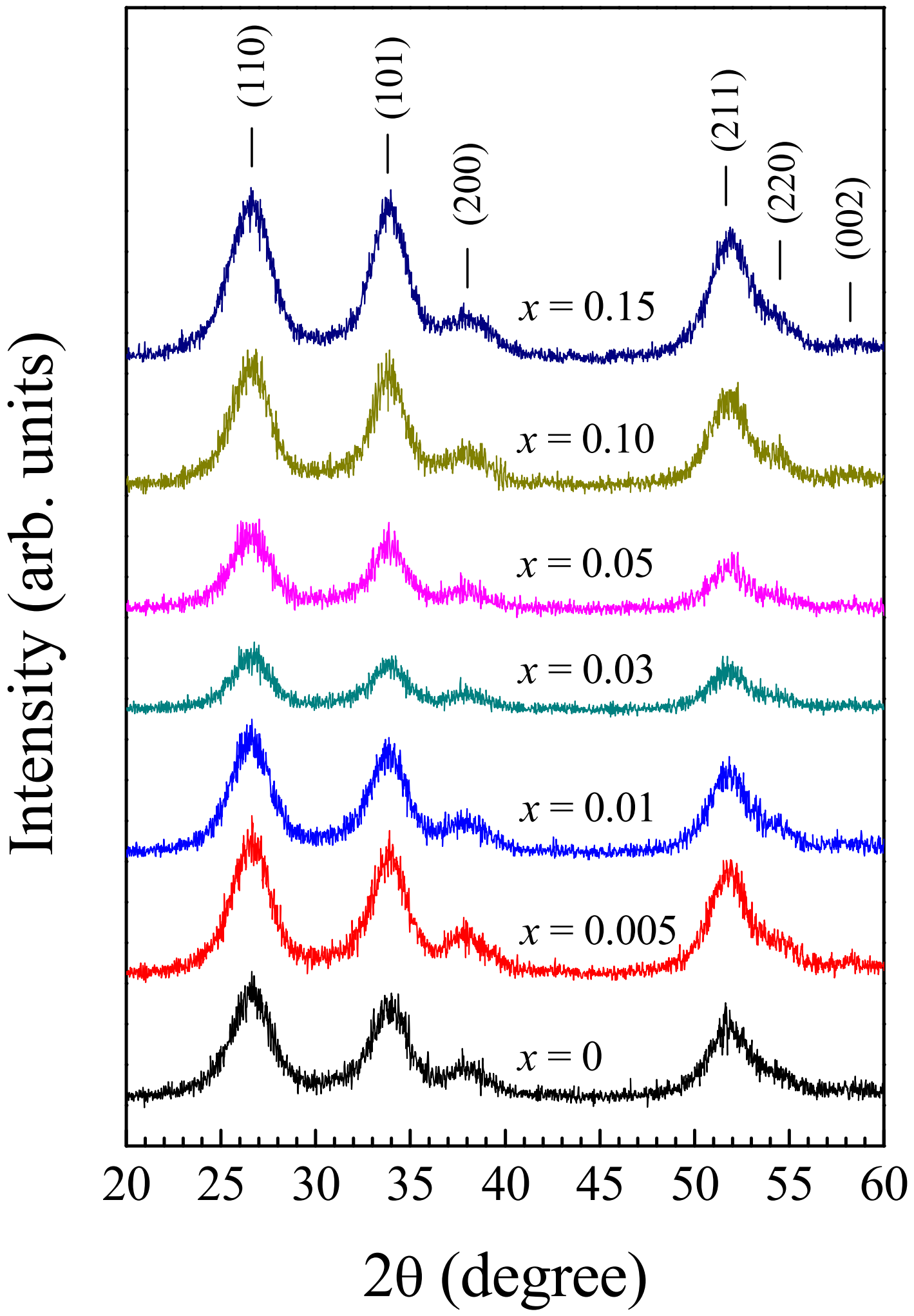

**Figure 2**

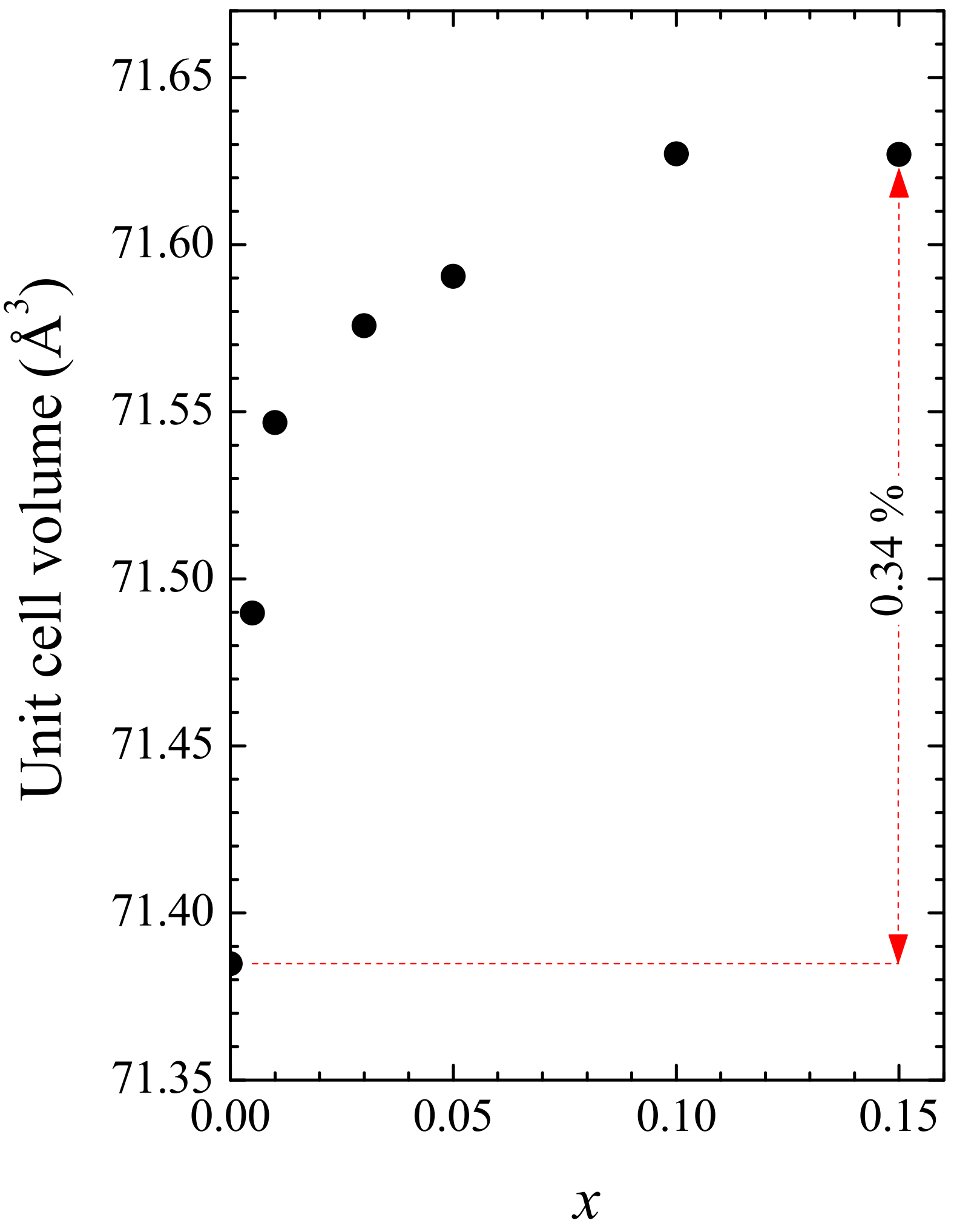

**Figure 3**

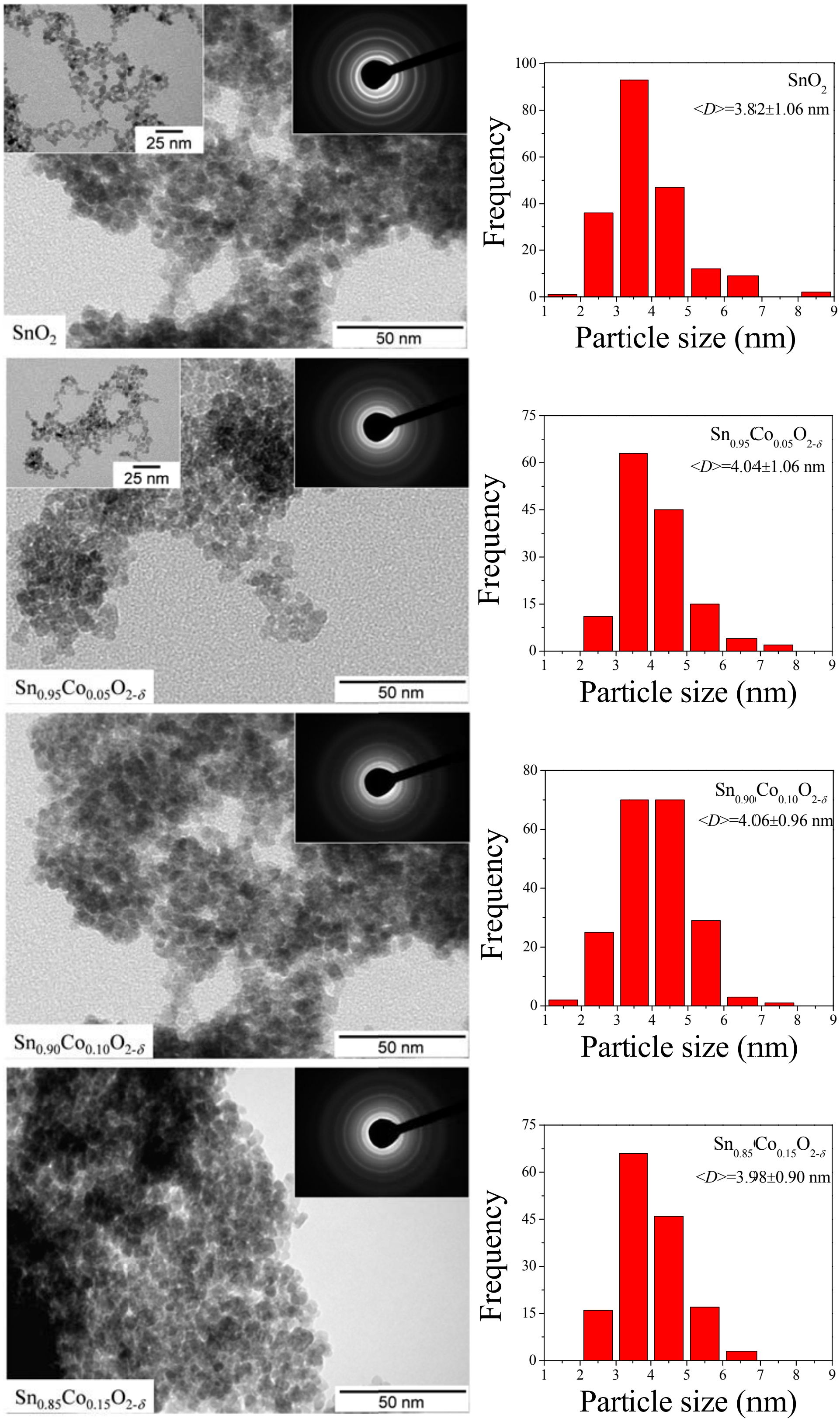

**Figure 4**

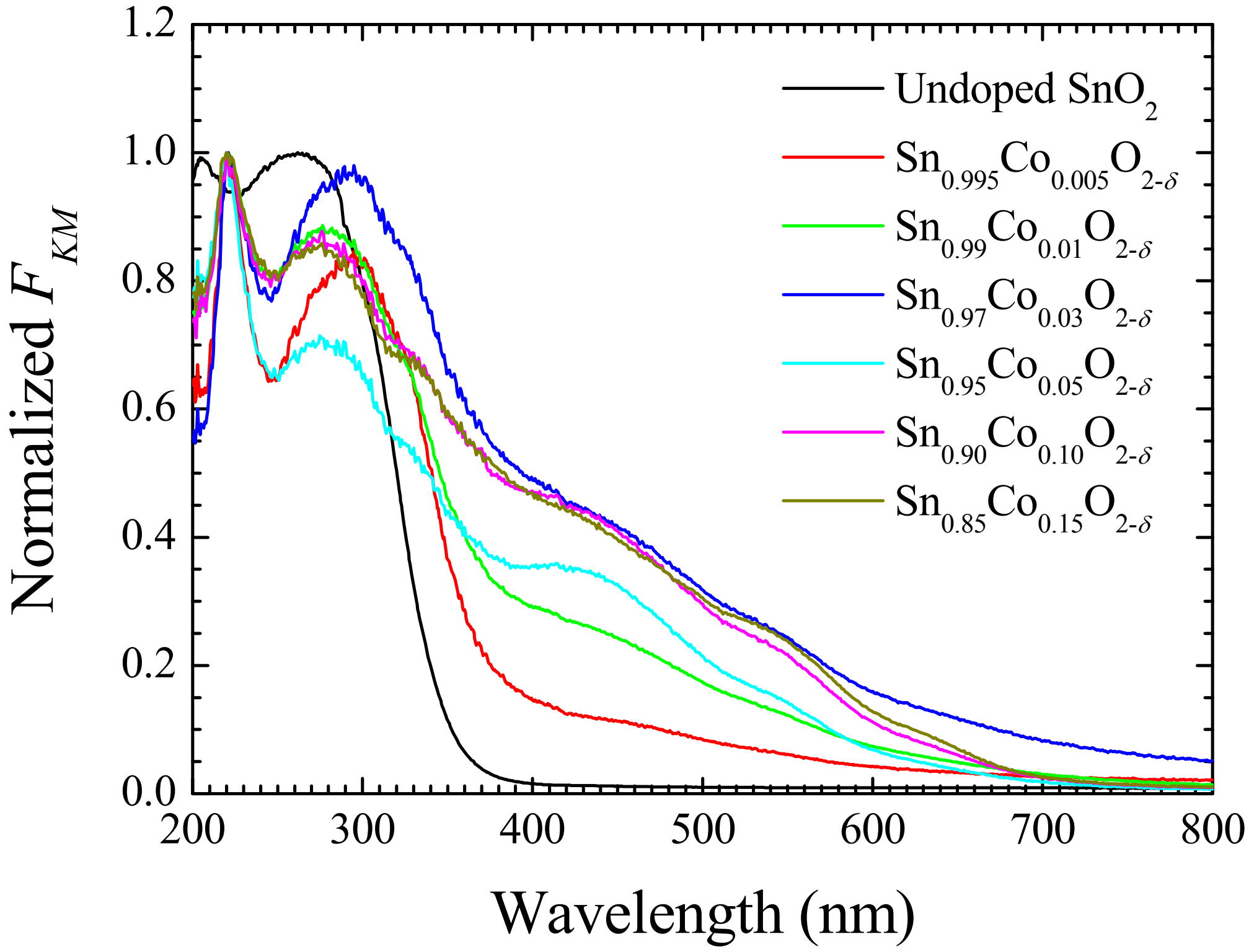

**Figure 5**

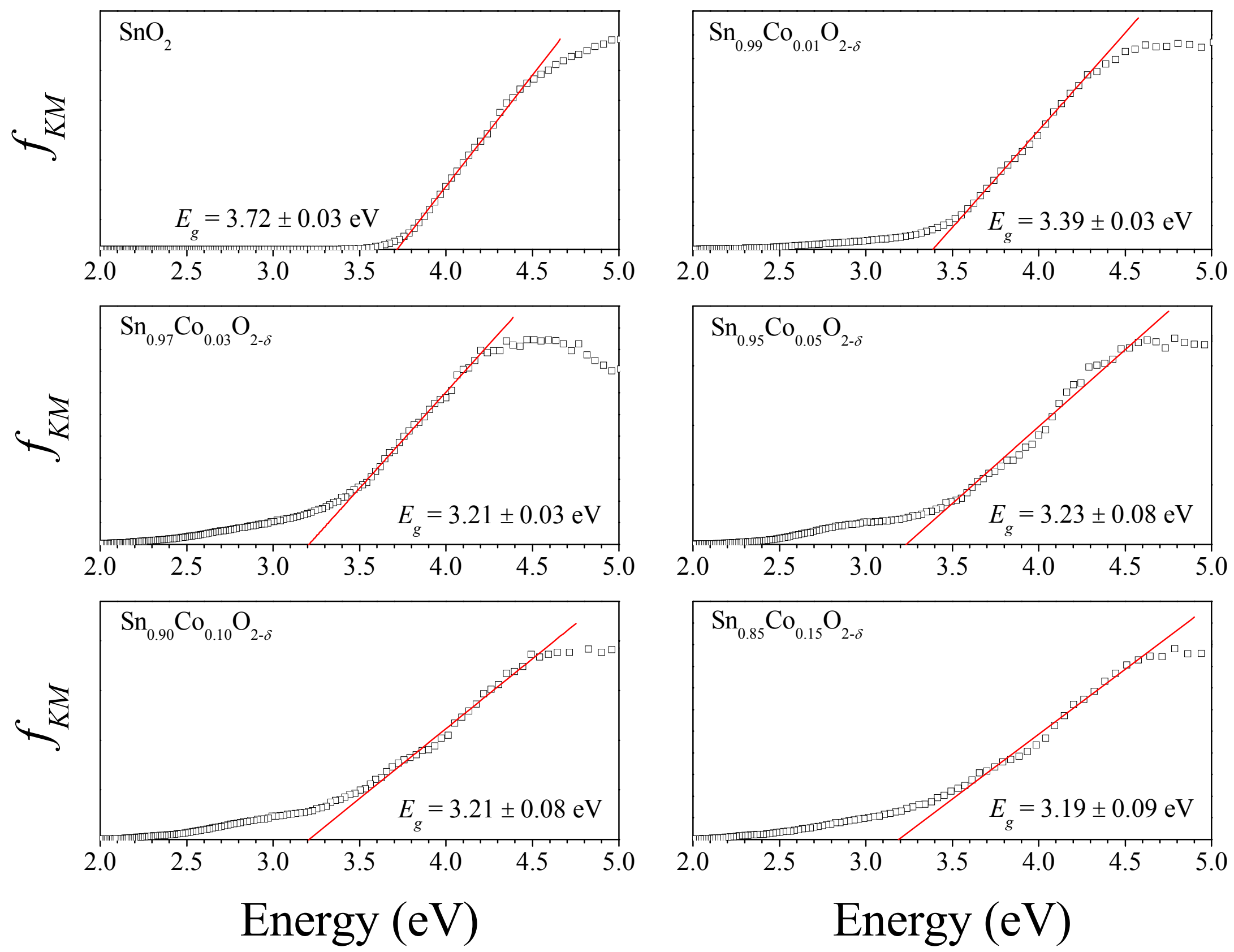
SnO2
$E_g$ = 3.72 ± 0.03 eV
Sn0.99Co0.01O2-δ
$E_g$ = 3.39 ± 0.03 eV
Sn0.97Co0.03O2-δ
$E_g$ = 3.21 ± 0.03 eV
Sn0.95Co0.05O2-δ
$E_g$ = 3.23 ± 0.08 eV
Sn0.90Co0.10O2-δ
$E_g$ = 3.21 ± 0.08 eV
Sn0.85Co0.15O2-δ
$E_g$ = 3.19 ± 0.09 eV
$f_{KM}$
2.0
2.5
3.0
3.5
4.0
4.5
5.0
Energy (eV)

**Figure 6**

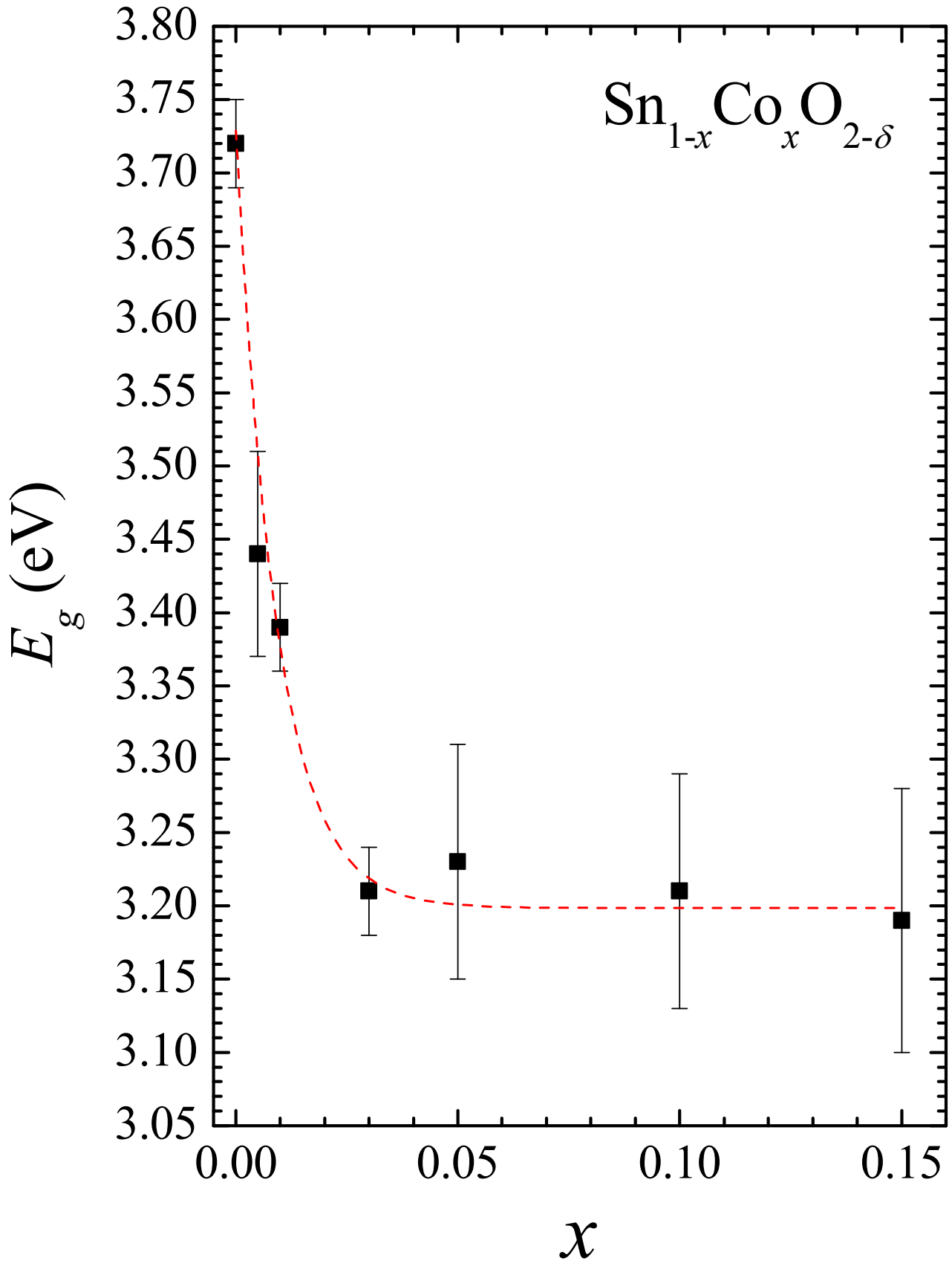

**Figure 7**

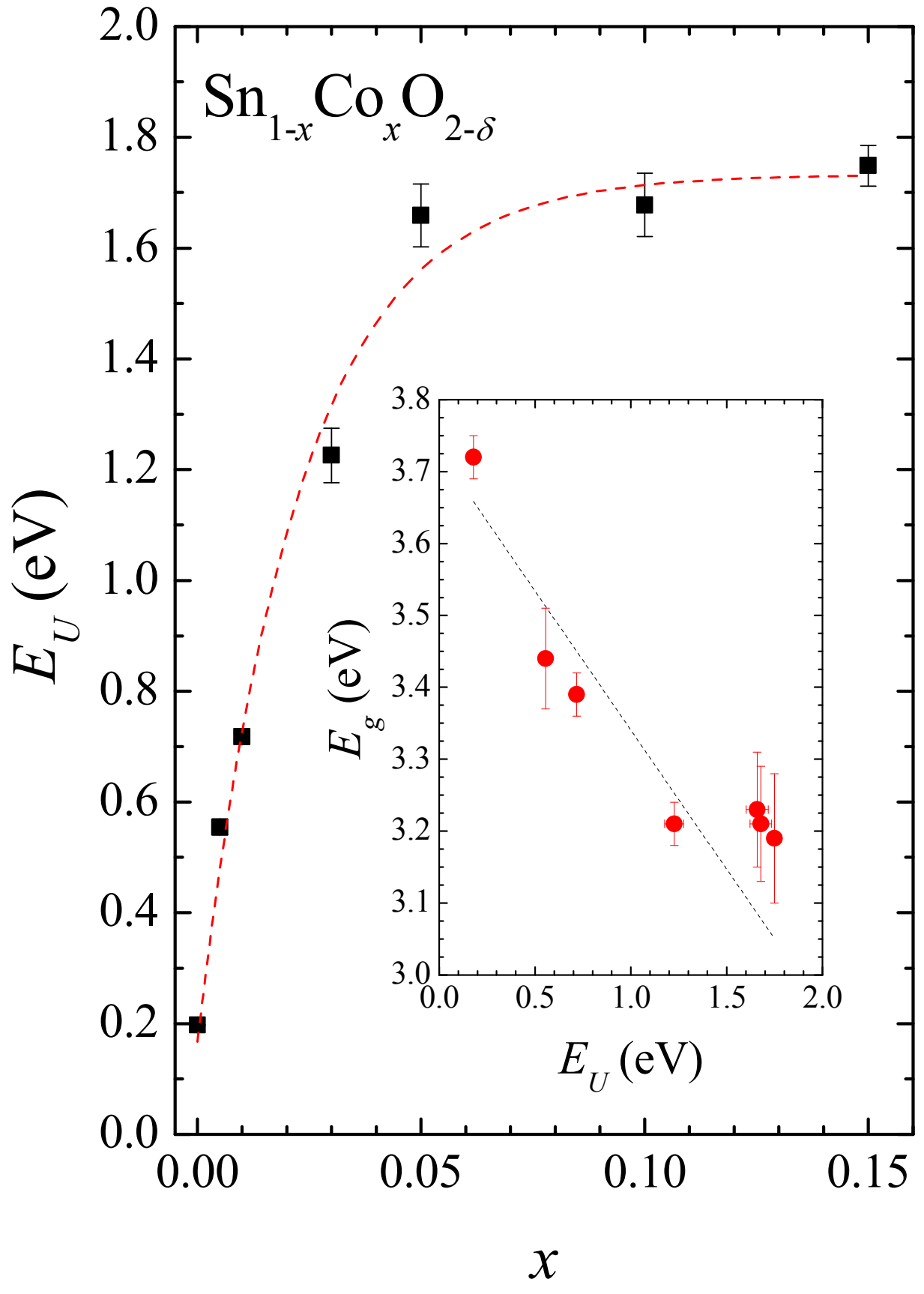

**Figure 8**

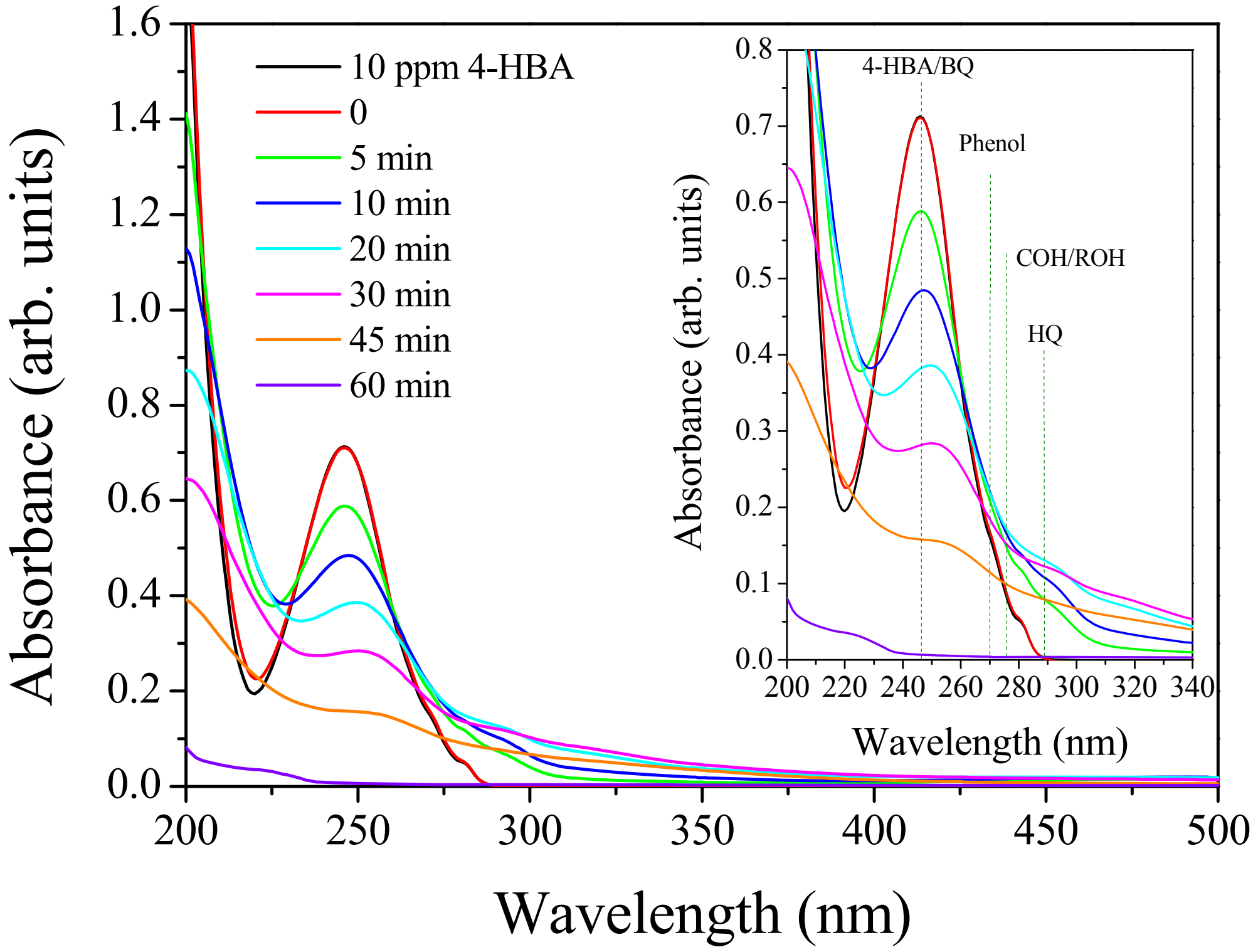

**Figure 9**

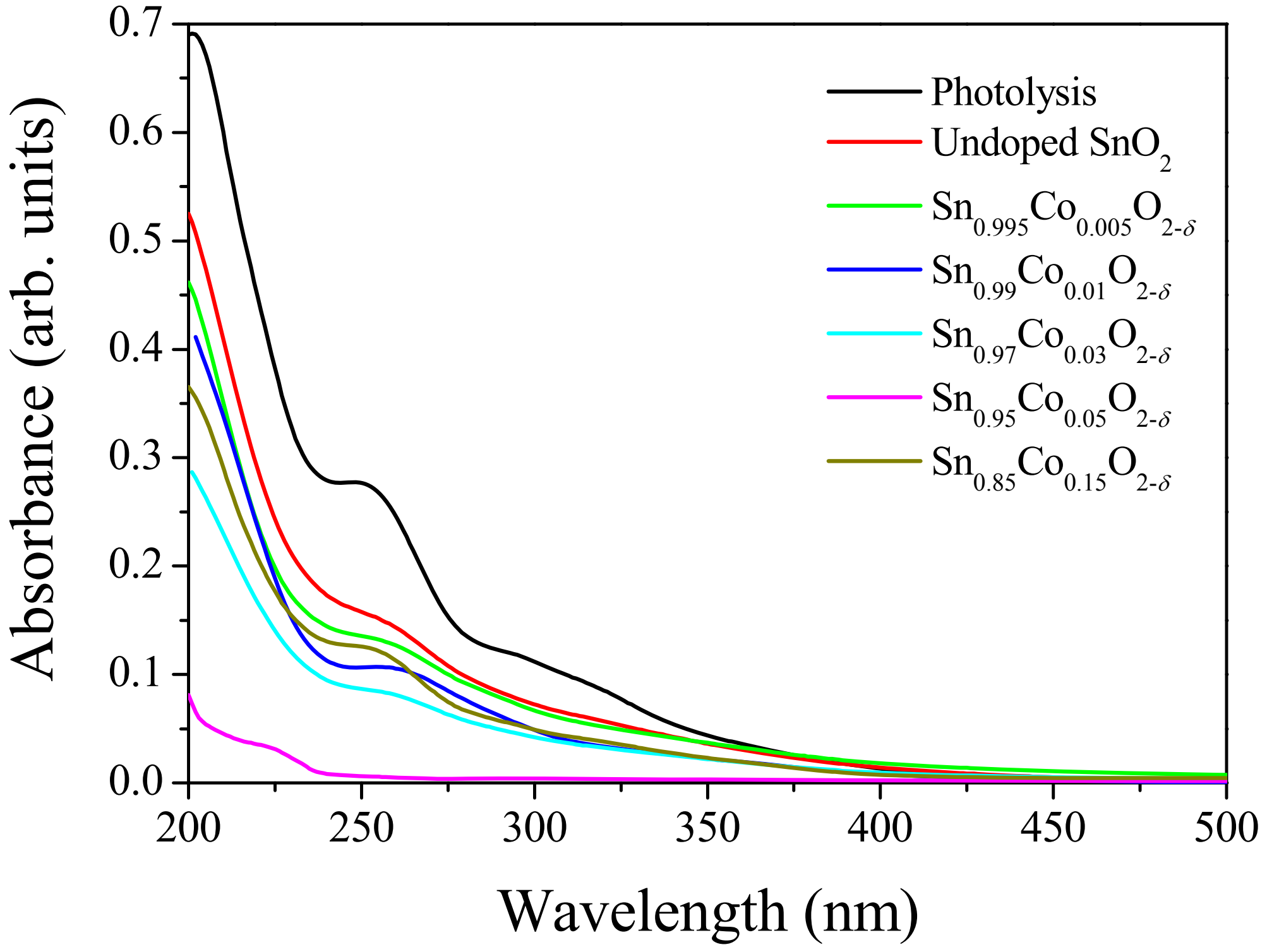

**Figure 10**

.